\documentclass[fleqn,usenatbib]{mnras}

\usepackage[T1]{fontenc}

\DeclareRobustCommand{\VAN}[3]{#2}
\let\VANthebibliography\thebibliography
\def\thebibliography{\DeclareRobustCommand{\VAN}[3]{##3}\VANthebibliography}

\usepackage{ragged2e}
\usepackage{color}

\usepackage{graphicx}	
\usepackage{amsmath}	

\usepackage{tikz}
\usepackage{epstopdf}

\newcommand{\art}{ART-XC}
\newcommand{\srg}{{\it SRG}}

\def\deg{\hbox{$^\circ$}}
\def\arcmin{\hbox{$^\prime$}}
\def\arcsec{\hbox{$^{\prime\prime}$}}

\usepackage[normalem]{ulem}

\def\deg{\hbox{$^\circ$}}
\def\arcmin{\hbox{$^\prime$}}
\def\arcsec{\hbox{$^{\prime\prime}$}}

\long\def\***#1{\textcolor{blue}{\textbf{\sffamily ***#1***}}}

\def\Swift{\emph{Swift}}
\def\Nustar{\emph{NuSTAR}}

\def\deg{\ensuremath{^\circ}}
\usepackage{newtxtext,newtxmath}

\title[Maximum likelihood X-ray source detection]{{\bfseries \itshape SRG}/\art\ Galactic Bulge deep survey. I. Maximum likelihood source detection algorithm for X-ray surveys}

\author[Semena et al.]{%
A. Semena$^{1}$,\thanks{E-mail: san@iki.rssi.ru (IKI)}
A. Vikhlinin$^{2,1}$,
I. Mereminskiy$^{1}$,
A. Lutovinov$^{1}$,
A. Tkachenko$^{1}$,\newauthor 
I. Lapshov$^{1}$,
R. Burenin$^1$\\
\vspace*{-0.5\baselineskip}\\
$^{1}$Space Research Institute RAS, 84/32 Profsoyuznaya str, Moscow 117997, Russia\\
$^{2}$Harvard-Smithsonian Center for Astrophysics, 60 Garden Street, Cambridge, MA 02138, USA\\
}

\date{Accepted XXX. Received YYY; in original form ZZZ}

\pubyear{2023}
\begin{document}
\label{firstpage}
\pagerange{\pageref{firstpage}--\pageref{lastpage}}
\maketitle

\begin{abstract}
We describe an X-ray source detection method entirely based on the maximum likelihood analysis, in application to observations with the ART-XC telescope onboard the \emph{Spectrum Roentgen Gamma} observatory. The method optimally combines the data taken at different conditions, a situation commonly found in scanning surveys or mosaic observations with a telescope with a significant off-axis PSF distortion. The method can be naturally extended to include additional information from the X-ray photon energies, detector grades, etc. The likelihood-based source detection naturally results in a stable and uniform definition of detection thresholds under different observing conditions (PSF, background level). This greatly simplifies the statistical calibration of the survey needed to, e.g., obtain the $\log N - \log S$ distribution of detected sources or their luminosity function. The method can be applied to the data from any imaging X-ray telescope.
\end{abstract}
 
\section{Introduction} 
\label{sec:intro}
 
The Mikhail Pavlinsky \art\ telescope is one of the two X-ray telescopes onboard the Spectrum Roentgen Gamma observatory \cite{srg}. The peak of its effective area is reached in 6--11~keV energy band. Its primary science objective is detection of the Galactic and extragalactic source populations in the hard band \citep{2019ExA....48..233P}. The most frequently used mode of ART-XC operations is wide-field surveys done with scanning observations. In this paper we describe the source detection algorithm developed specifically for \art\ surveys, but which also is widely applicable to any imaging X-ray telescope. 

The detailed description of the \art\ telescope is given in \citet{2021A&A...650A..42P}. Here, we briefly list its characteristics pertaining to our analysis. \art\ consists of seven co-aligned grazing incidence Wolter I \citep{1952AnP...445...94W} telescopes. The full operating energy range is $4-35$~keV, but the telescope is most sensitive in the $4-12$~keV band. The field of view (FoV thereafter) is approximately circular with a radius of 18\arcmin. The telescope uses CdTe semiconductor double-sided strip detectors with a spatial resolution of $45$\arcsec\ \citep{2014SPIE.9144E..13L}. The particle-induced background is a dominating component and its distribution on the detector is nearly flat. It is highly stable due to SRG operating in the Earth-Sun second Lagrange point, except for periods of Solar flares \citep{2021A&A...650A..42P}. 

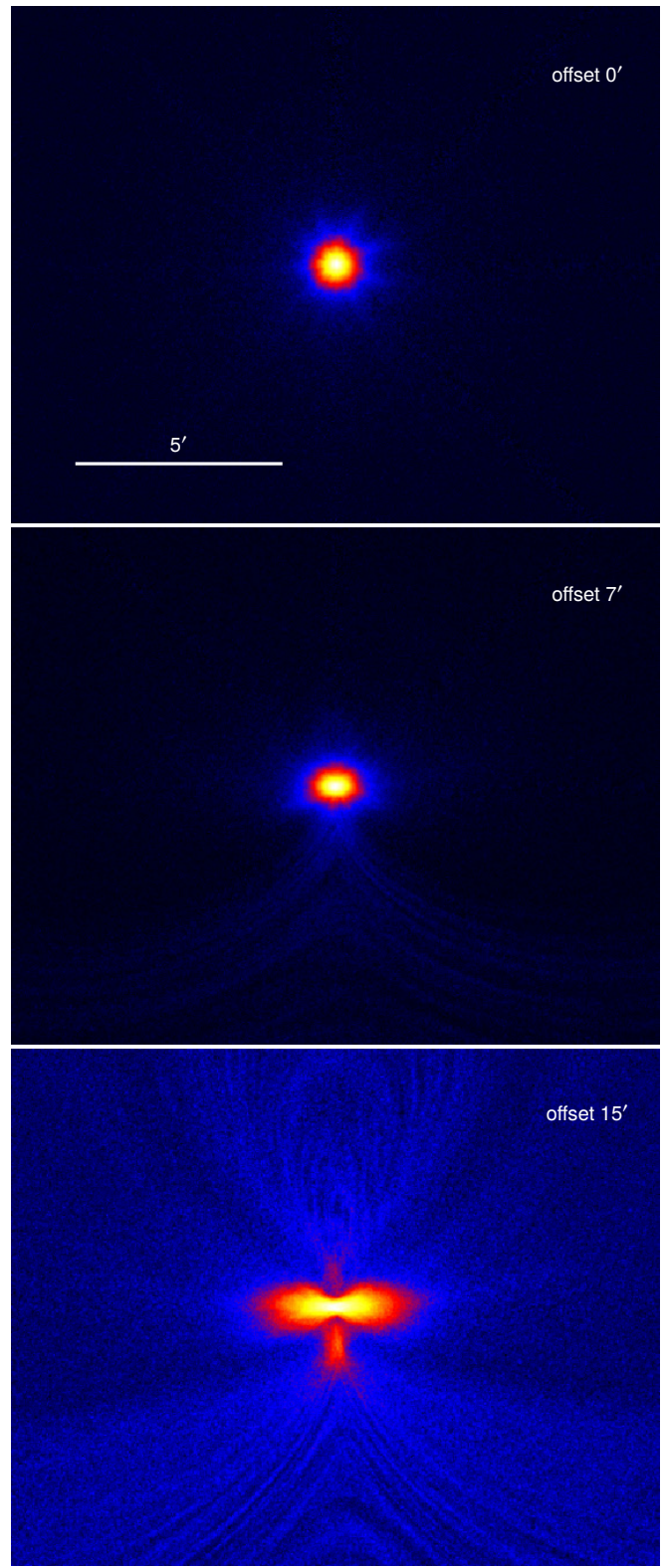
\begin{figure}
\vspace*{1mm}
    \centering
\begin{tikzpicture}[scale=\columnwidth/1cm] 
\node[anchor=south west,inner sep=0] (image) at (0,0) {\includegraphics[width=\columnwidth]{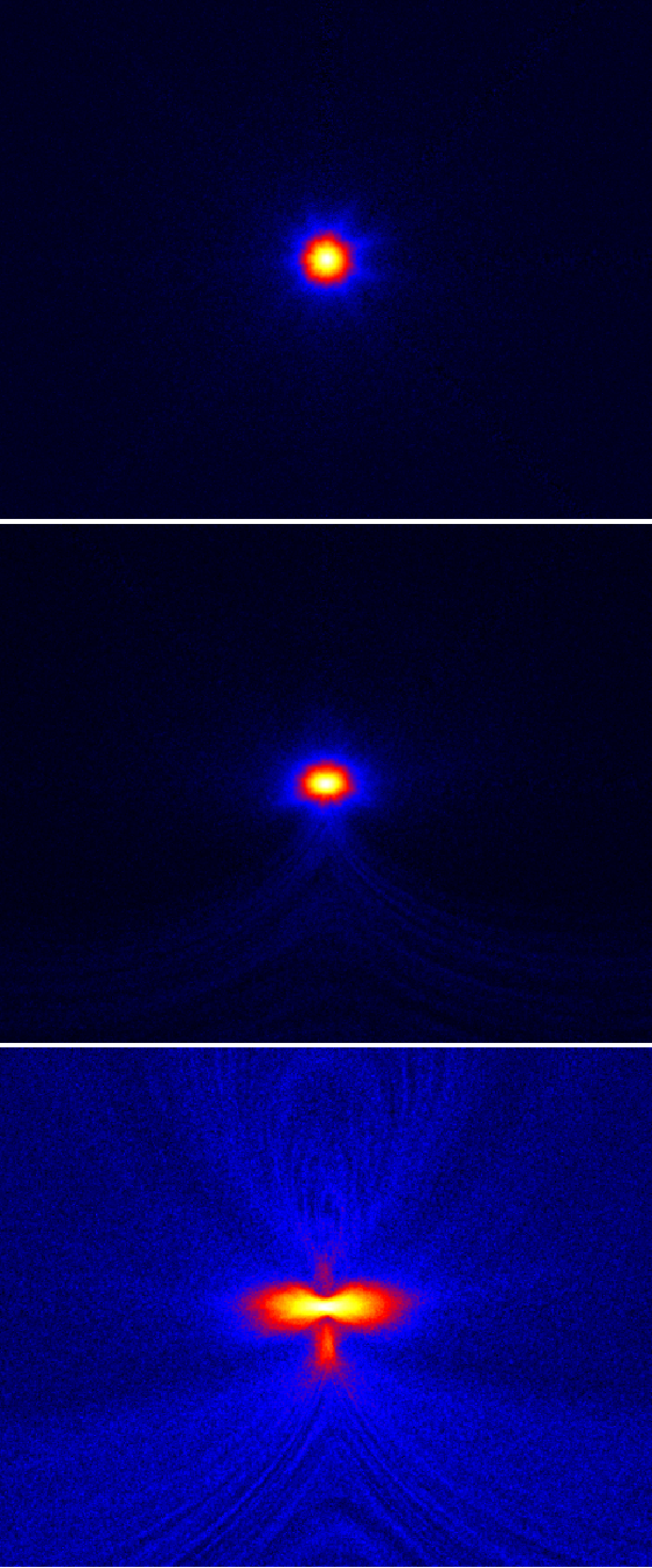}};
\draw[white, very thick] (0.1, 1.7) -- (0.41914, 1.7); 
\node[text=white] at (0.26, 1.73) {\textsf{5}\arcmin};
\node[text=white] at (0.89, 2.3) {\textsf{offset 0}\arcmin};
\node[text=white] at (0.89, 1.5) {\textsf{offset 7}\arcmin};
\node[text=white] at (0.89, 0.7) {\textsf{offset 15}\arcmin};
\end{tikzpicture}

    \caption{The Point Spread Function (PSF) of the \art\ mirrors measured in the ground tests. The telescope has been intentionally slightly defocused by 7mm to improve the angular resolution off-axis \citep[for details, see ][]{2019ExA....48..233P}. As a result, the PSF remains compact ($\mathrm{HPD}=30\arcsec$) and mostly circular within $\approx9$\arcmin\ from the optical axis. Outside this radius, the PSF width quickly degrades and its shape becomes highly distorted (\emph{bottom panel}). Much of these distortions are due to single reflections in the \art\ mirror system \citep[cf.\ Fig.9 in][]{2019ExA....48..233P}.}
    
    \label{fig:IPSF}
\end{figure}

The main limiting factor for the ART-XC source detection is a relatively wide point spread function (PSF thereafter) and a relatively high particle-induced background. The net effect of these two factor is that the source photons are spread over a large area and the signal should be detected against a significant level of Poisson noise. An additional complication for classical source detection methods is a strong variation of the ART-XC response (the PSF, mirror vignetting, background) within the field of view. Examples of the ART-XC PSF measured during ground tests \citep{2017ExA....44..147K} are shown in Fig.\ref{fig:IPSF}. The PSF size and shape strongly changes as the source is moved away from the optical axis. It is circular with a $\sim30$\arcsec\ HPD in the central 9\arcmin\ but then widens to $\sim 2\arcmin$ and becomes strongly elongated towards the FoV edge \citep[see Fig.13 in][for more information]{2021A&A...650A..42P}. During the scans, the source spends $\approx$ 25\% of the time in the central FoV zone with a compact PSF. Approximately half of the source photons are collected in this zone. The other half of the source photons are collected when it is in the outer regions of the FoV with the poor PSF. Each source is therefore observed under very heterogeneous conditions, and there is no clear boundary one can use to simply filter out  observing periods with poor data quality. 

The optimal source detection should take all these changes into account instead of operating on a single, combined image as is typically done for pointed observations \citep[see, e.g., for a recent review][]{2012MNRAS.422.1674M}. Our detection method does just that by combining the likelihoods computed individually at the time of registration of each photon, as well as by using additional information from the photon energy and detector grades. The basic information we use to identify sources is the map of the likelihood ratio test, which for each sky location provides a measure of probability that the observed photon distribution can be attributed entirely to the background. A high value of the likelihood ratio at a given sky location corresponds a low probability for the ``background-only'' hypothesis, and hence implies an existence of a source. This approach represents a statistically optimal test for detection or rejection of sources \citep{1933RSPTA.231..289N}.

The fully likelihood-based source detection approach has indeed been historically used for analyzing data from the $\gamma$-ray observatories, which are characterized by a complex detector response and where one is often forced to work at lower levels of statistical significance \citep{1996ApJ...461..396M,1999ApJS..123...79H, 2010ApJS..188..405A}. For the \emph{imaging} X-ray telescopes, the maximum-likelihood analysis has been applied in a more limited manner. 
It has been used for modeling of detected sources in the analysis pipelines, e.g., for \emph{ROSAT}~\citep{1988ESOC...28..177C}, \Swift~\citep{2020ApJS..247...54E}, \emph{XMM-Newton}~\citep{2004ASPC..314..759G}, and the
\emph{Chandra} Source Catalog~\citep{2010ApJS..189...37E}. However, in
all prior implementations, the maximum-likelihood fitting was the
second stage of a two-step procedure in which the candidate sources
were first detected by much cruder methods. For example, the \Swift\ source
catalog \citep{2020ApJS..247...54E} uses the sliding box detection
while the \emph{Chandra} Source Catalog approach is equivalent to
using an approximately Gaussian detection filter. The prime reason for using these simplified methods is a high computational cost of performing the likelihood-based analysis at all locations. We solved this issue by 
heavily optimizing the process of maximizing the likelihood function.
As a result, \emph{our detection
  technique uses the likelihood-based approach at the crucial step of
  initial source detection.} This naturally leads to other advantages
such as the statistically optimal combination of data taken at
non-uniform conditions (background, PSF, etc.) or even a combination
of data from different telescopes.

A conceptually similar algorithm has been described previously by \citet{2018AJ....155..169O} and \citet{LYNXREPORT}, and applied to obtain the catalog of a point sources from the first year of the ART-XC all-sky survey \citep{art_xc_allsky_survey}. The critical improvement introduced in our approach is a full likelihood analysis at each sky location that not only established existence of a source but also returns its best-fit flux. As a result, we automatically measure the fluxes of detected sources. Our method also leads to a stable and uniform definition of detection thresholds and estimation of the number of false detections. 

While the bulk of our work is generic and applicable to the X-ray data from any imaging X-ray telescope (\emph{Chandra}, \emph{XMM-Newton}, \Swift, \Nustar, \emph{eROSITA}), there are several ART-XC specific issues that we had to address. They include a procedure for estimating the local backgrounds and reducing the illumination from bright off-axis sources during the survey scans. Our solutions to these problems are instructive and are given below.

The paper is organized as follows. In section \ref{sec:detection}, we describe the detection method. The mathematical basics are given in \S\ref{sec:likelihood_ratio}; the iterative maximum likelihood fit and flux measurement procedure are described in \S\ref{sec:flux}; our treatment of the detection thresholds in terms of the likelihood ratio is described in \S\ref{sec:detection_threshold}, estimation of the positional and flux count rate uncertainty is described in section \ref{sec:confidece_intervals}. In \S\ref{sec:illumination}, we discuss how the illumination from bright sources is suppressed. The background estimation procedure for the ART-XC scans is detailed in \S\ref{sec:background_model}.

\section{Source detection}
\label{sec:detection}

The \art\ detector background is very stable, excluding periods of Solar flares \citep{2021A&A...650A..42P} but relatively high ($\sim0.4$~cts~s$^{-1}$ per the 1020~arcmin$^2$ field of view in 4--12~keV energy band in each telescope). The expected depth of the typical \art\ scientific observation programs (a few ksec total exposure) is a $\mathrm{few}\times10^{-13}$~erg~s~cm$^{-2}$ \citep{2019ExA....48..233P}. The background contribution in the PSF beam is non-negligible for sources in this flux range $\sim55$ background  vs.\ $\sim 22$ sources counts detected within the 1~arcmin PSF beam for 3ks exposure.
Moreover, the majority of the \art\ observations are done in the ``scanning'' mode, which results in each source being exposed roughly uniformly within the field of view. This creates a wide mix of significantly varying PSF (see Fig.\ref{fig:IPSF}), and also the source-to-background ratios due to non-uniform vignetting. Under these conditions, only a highly optimal source detection method allows us to address the main scientific objectives of \art\ of surveying  large regions in the sky in the hard X-ray band \citep{2021A&A...650A..42P}.

As we discussed above, the likelihood-based analysis is the theoretically most optimal method for detecting faint sources. It is also the optimal approach for combining non-uniform data sets \citep{lehmann2005testing}, such as exposures of the same source with different PSF in our case. Therefore, we base our method on the likelihood function analysis.

The expected source density at a flux limit of $10^{-13}$~erg~s~cm$^{-2}$ in the 4--12~keV band is 5.8~deg$^{-2}$ in the extragalactic sky \citep{2007A&A...463...79G}  and $\approx 15$~deg$^{-2}$ in the Galactic Ridge \citep{2004MNRAS.351...31H}. This translates to 0.003 and 0.01 sources per the PSF area, respectively. Therefore, \art\ is unaffected by source confusion except for a small area around the Galactic Center, and its typical detection regime is that of finding isolated point sources. We provide a rigorous treatment of the likelihood-based source detection for such a case. In section 2.1 below, we start the description of our method with the discussion of the likelihood calculation for a source in a given sky position.

\subsection{Likelihood ratio test for existence of a source}
\label{sec:likelihood_ratio}

The basic observable for an X-ray telescope is a set of events with a known arrival time, position on the detector, and energy. For a source with a given flux and sky position, we have a model for the count rate at this detector location and arrival time (see discussion near eq.~\ref{eq:Npix} below). Following the approach of \citet{1979ApJ...228..939C}, we can write the Poisson probability to observe a set of events in a single pixel, given a model, as follows: 
\begin{equation}
    \label{eq:poisson_prob_base}
    P = \prod_t\frac{1}{n!}\exp(-\mu\,\delta t) (\mu\, \delta t)^n,
\end{equation}
where the observation is split into very short intervals $\delta t$ such that at most one event is detected in each interval; $\mu$~$(=\mu(t)$) is the model count rate; and $n=1$ if the event is detected and 0 if not. With this fine binning, eq.~\ref{eq:poisson_prob_base} can be rewritten as
\begin{equation}
    \label{eq:1pix}
    P = \exp\left(-\int \mu\,dt\right) \prod_i \mu_i\delta t,
\end{equation}
where the product is over the detected set of events.
For a set of independent pixels, the probability is a product of probabilities given by eq.~\ref{eq:1pix}. It can be simplified as follows:
\begin{equation}
   \label{eq:Npix}
    P = \exp{\left( -\sum_{\text{pix}}\int \mu dt \right)} \prod_i \mu_i\delta t.
\end{equation}
The model count rate $\mu$ is, generally, a sum of the source ($r$) and background ($b$) models: $\mu=r+b$, and so eq.~\ref{eq:Npix} can be written as
\begin{equation}
   \label{eq:poisson_prob}
    P = \exp{\left( -\sum_{\mathrm{pix}}\int (r + b) dt \right)} \prod_i (r_i + b_i)\delta t. 
\end{equation}

Let us now consider how the source model, $r$, can be factored (the background model is discussed in \S\,\ref{sec:background_model} below). Assuming that the source flux is constant over the observing period, we can write $r = R s$, where $R$ is the count rate for a source located at the optical axis, and $s$ can be written as
\begin{equation}
\label{eq:source:model}
    s(\xi,\eta,E) = V(\xi,\eta,E) \int \mathrm{PSF}(x-\xi,y-\eta,\alpha,\beta,E)\,dx\,dy
\end{equation}
where $\mathrm{PSF}(\Delta x,\Delta y,\xi,\eta,E)$ is the position- and energy-dependent point spread function, i.e. the probability of photon scattering to location $(\Delta x,\Delta y)$ relative to the nominal source location within the field of view, $(\xi,\eta)$; $V(\xi,\eta,E)$ is the vignetting factor at the nominal source location; and the integral is over one pixel, necessary if the pixel and PSF sizes are comparable as is the case for ART-XC. The time dependence in eq.~\ref{eq:poisson_prob} is encoded in eq.~\ref{eq:source:model} through the time dependence of $\xi$ and $\eta$ that describes how the source moves through the field of view during scans.

From the probability of the observed data given the model (eq.~\ref{eq:poisson_prob}, \ref{eq:source:model}), we can now construct the classical likelihood ratio test for rejecting a case that the observed photons are entirely produced by the background \citep{1933RSPTA.231..289N}:
\begin{equation}
\label{eq:orig_lkl_ratio}
    \Delta \ln{L} = \ln{\frac{\max_R P(R)}{P(0)}} = \max_R \left[\sum_i \ln{\frac{R s_i + b_i}{b_i}} - e R\right],
\end{equation}
where (cf.\ eq.~\ref{eq:source:model})
\begin{equation}
    \label{eq:vign_exp}
    e = \sum_{\mathrm{pix}} \int s dt = \int V(\xi,\eta,E)\,dt
\end{equation}
The physical meaning of $e$ is the exposure-averaged vignetting correction.

Note that while formally the summation in eq.~\ref{eq:orig_lkl_ratio} is over all photons, in practice only the photons sufficiently close to the given location contribute significantly. For photons with large $\Delta x$, $\Delta y$, $s\rightarrow 0$, and \begin{math}\ln{\big({(Rs+b)/b}\big)}\rightarrow 0\end{math}, also. Therefore we set the ART-XC PSF model to zero outside of the $5'$ radius. This aperture contains 95\% of the total source flux near the FoV edge, and nearly 100\% of that on-axis. The photons that enter the likelihood calculations are then also selected within $5'$ of the given position. This area contains $0.04 (f/10^{-13} \mathrm{erg}\,\mathrm{s}^{-1}\mathrm{cm}^{-2})^{-1.5}$ sources on average \citep{2007A&A...463...79G}, hence the assumption of a single source implicitly used in our derivation is valid.

The value of $R$ that maximizes the likelihood ratio in eq.~\ref{eq:orig_lkl_ratio} can be found, as usual, by solving the equation 
\begin{equation}
\label{eq:count_rate_condition}
\frac{d\,\ln P(R)}{dR} = 0.
\end{equation}
After taking derivatives, eq.~\ref{eq:count_rate_condition} can be rewritten as follows:
\begin{equation}
    \label{eq:rate_condition_explicit}
    \sum \frac{s_i}{R s_i + b_i} = e
\end{equation}

\begin{figure*}
    \centering
    \includegraphics[width=0.49\textwidth]{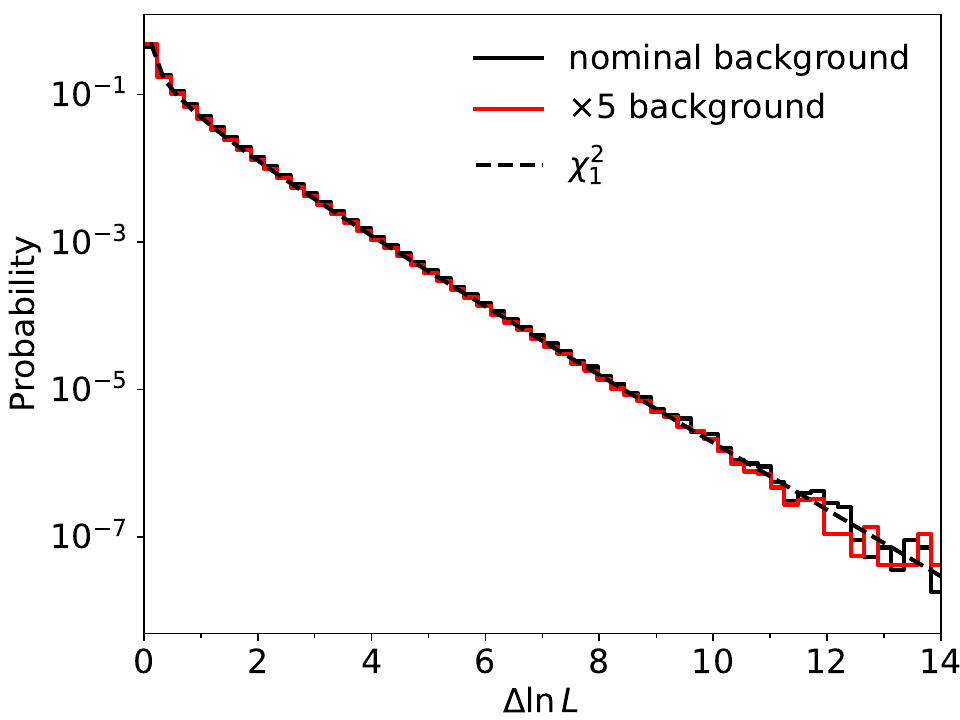}\hfill
    \includegraphics[width=0.49\textwidth]{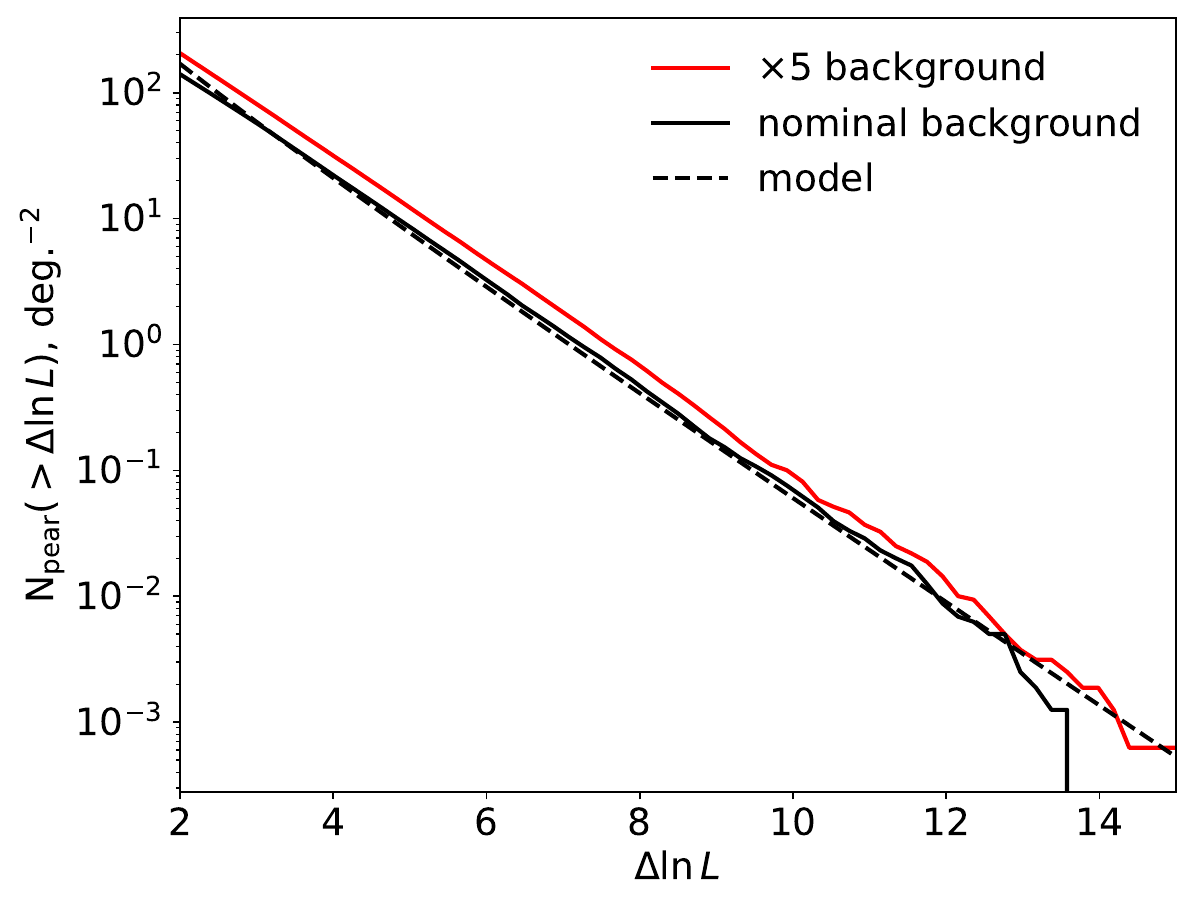}
    
    \caption{\emph{Left:} The distribution of the $\Delta \ln{L}$ values in a simulated \art\ scan observation of an empty region. As expected, the $\chi^2_1$ distribution provides an excellent approximation to the observed probability density function. We observe an identical probability density function for a background level five times the nominal, also as expected (see text). \emph{Right:} 
    The peak value distribution of the local maxima in the simulated $\Delta \ln{L}$ map. This distribution represents the false detection probability. Note that it is not equivalent to the distribution of all values in the $\Delta \ln{L}$ map, shown in the left panel. The distribution of peak values remains only weakly dependent on the background level, with number of false detections changing by 20\% for a factor of 5 higher background level. The dashed line represents the model, which approximates the expected cumulative number of false detections (see discussion \S\ref{sec:detection_threshold}).}
    
    \label{fig:dist}
\end{figure*}

Many existing source detection packages \citep[see e.g. XMM SAS][]{2004ASPC..314..759G} implement an equivalent of the procedure described above, for a subset of trial source locations, identified by other means --- usually, via simplified source detection methods such as sliding box or wavelet transform (e.g. \citet{1999A&A...349..389V, 2022A&A...661A...1B}, check also a more comprehensive background subtraction technique which can advance such initial analysis \citet{2022MNRAS.515.5185E}). Clearly, this first step leads to an overall sub-optimal detection , although in the regime of very high background levels the matched filter becomes identical to the likelihood ratio test \citep{1968IJTP....1...37H, 2018A&A...616A..25V}. The prime reason for using these simplified method is the absence of analytical solutions to eq.~\ref{eq:rate_condition_explicit} and a high computational cost of solving eq.~\ref{eq:rate_condition_explicit} numerically because in general it has no analytic solution. The source rate solution can be straightforwardly performed with a family of gradient-free root finding solvers \citep{zbMATH06193572}, which usually require more then 10--20 iterations to achieve a percent level of convergence. Note that each evaluation of eq.\ref{eq:rate_condition_explicit} involves a substantial amount of vector algebra, and if one aims to find the solution for $R$ at each location covered by a dataset, the computatonal costs accumulate. Below we describe a faster and highly accurate method that enables the implementation of the likelihood-based detection scheme for maximizing the likelihood on a desktop-class computer.

\subsection{Iterative calculation of the likelihood map}
\label{sec:flux}

We start with modifying eq.\ref{eq:rate_condition_explicit} as follows:
\begin{equation}
    \frac{1}{e}\;\sum \frac{s_i}{R s_i + b_i} = 1,
\end{equation}
\begin{equation}
    \label{eq:iter:scheme:0}
    R=\frac{1}{e}\;\sum \frac{R s_i}{R s_i + b_i}.
\end{equation}
This equation can be used as a basis for the iterative scheme,
\begin{equation}
    \label{eq:iter:scheme:1}
    R''=\frac{1}{e}\;\sum \frac{R' s_i}{R' s_i + b_i},
\end{equation}
where $R'$ and $R''$ are the previous and new iterations of $R$, respectively. Let us show that this iterative scheme does indeed converge. Let $R$ is the exact solution of eq.~\ref{eq:rate_condition_explicit} (and, equivalently,~\ref{eq:iter:scheme:0}), $\delta R'=R'-R$ and $\delta R'' = R'' - R$ are the previous and new deviations of iterations from that value. Then (cf.~eq.~\ref{eq:iter:scheme:0} and~\ref{eq:iter:scheme:1}),
\begin{equation}
\begin{split}
    \label{eq:iter:scheme:convergence}
    \delta R'' = \frac{1}{e}\sum \frac{R' s_i}{R' s_i + b_i} - \frac{1}{e}\sum \frac{R s_i}{R s_i + b_i} = \\ 
    \delta R' \times \frac{1}{eR}\sum \frac{R s_i}{R s_i + b_i}\times\frac{b_i}{(R' s_i + b_i)}.
\end{split}
\end{equation}
Now note that per eq.~\ref{eq:iter:scheme:0}, $(eR)^{-1}\sum (R s_i)/(R s_i + b_i)=1$ and therefore $\delta R''<\delta R'$ if $R'>0$ and thus the iterations always converge. In practice, $<20$ iterations are needed to obtain a solution for $R$ with a 1\% accuracy. We also checked that the convergence is largely independent on the initial value of $R'$.

Implementation of this method opens us a possibility of computing the map of the likelihood ratio over the entire observed region (all sky in the ART-XC case). Local maxima of a sufficient height in such a map correspond to real sources. Moreover, the value of $R$ obtained by solving eq.~\ref{eq:iter:scheme:0} is, automatically, the maximum likelihood estimate of the source countrate.

\subsection{Detection thresholds in terms of $\boldmath\Delta \ln L$}
\label{sec:detection_threshold}

Real sources correspond to local maxima in the likelihood ratio map. We need to set a reasonably high threshold to avoid accepting a large number of false sources, i.e. the source is detected if
\begin{equation}
    \Delta \ln{L} > \Delta \ln{L}_{\rm lim}.
\end{equation}
This definition of the detection threshold leads to a robust and stable estimation of the number of false detections that can be done nearly from the first principles.
Indeed, according to the Wilks theorem \citep{10.1214/aoms/1177732360}, the probability distribution of the quantity $2\,\Delta\ln{L}$ should approach to the distribution $\chi^2_1$ for the case when there really is no source in the sky ($R = 0$). This implies that that the distribution of $\Delta \ln{L}$ should be independent of the background level. A direct Monte Carlo simulation shown in Fig.~\ref{fig:dist}a demonstrates that indeed the distribution of $\Delta \ln{L}$ does not depend on the background and follows the $\chi^2_1$ law.

\begin{figure}
    \centering
    \includegraphics[width=\columnwidth]{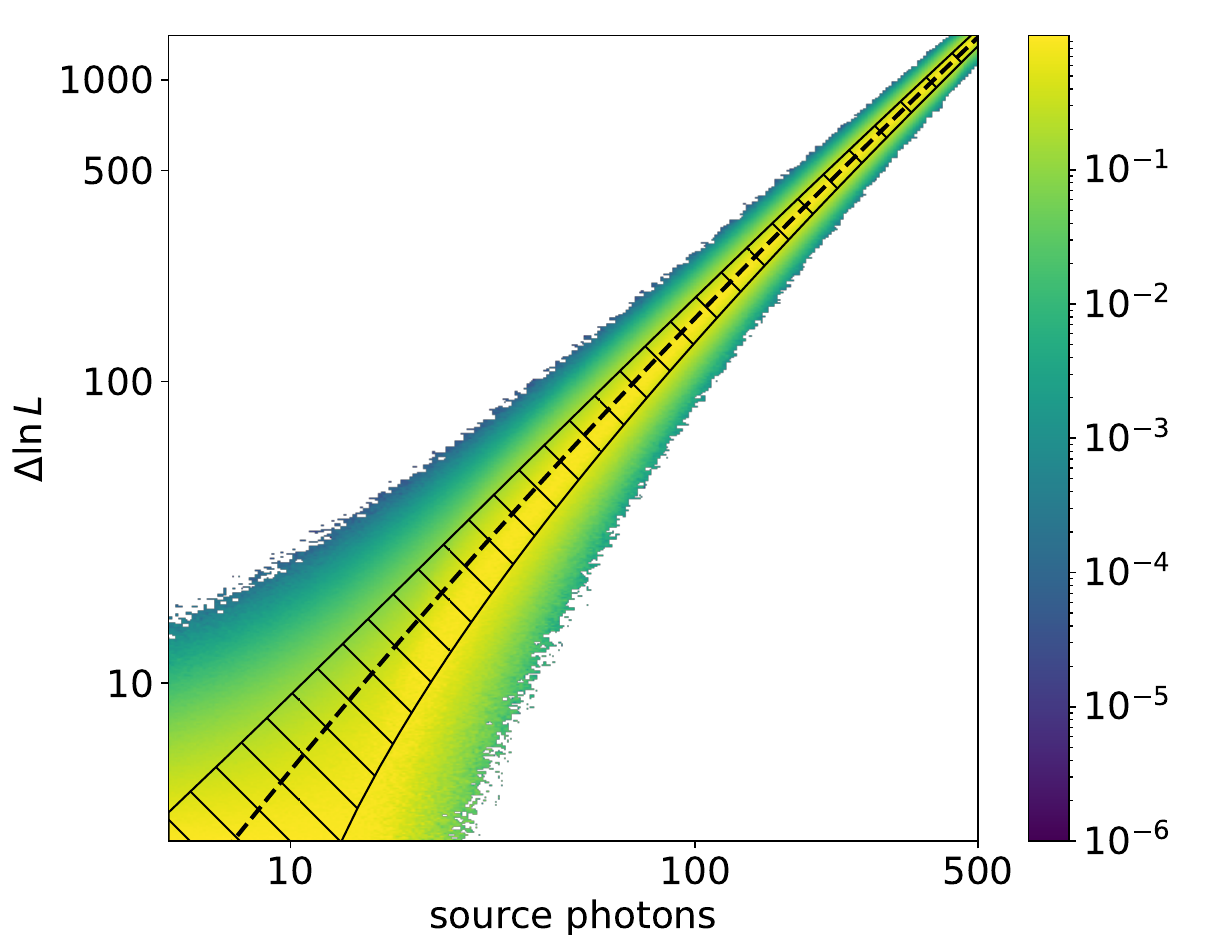}
    \caption{A relationship between the source flux and $\Delta \ln{L}$, obtained via Monte Carlo simulations of a typical \art\ scanning observation of a $20~$deg$^2$ region. The dashed line shows the mean $\Delta \ln{L}$. 
    For bright sources, the distribution of $\Delta \ln{L}$ around the mean is close to the normal distribution due to the central limiting theorem because it is the sum of large number independent random variables (eq.\ref{eq:orig_lkl_ratio}), while for faint sources it should approach the $\chi^2_1$ (\S\ref{sec:detection_threshold}). The the hatched area corresponds to the $\pm1\sigma$ band around the mean $\Delta \ln{L}$.}
    \label{fig:delta_ln_l_distribution}
\end{figure}

Note, however, that to characterize the probability of the false detection we need the distribution of the peak heights of the \emph{local maxima}, not all $\Delta\ln L$ values across the field. The $\Delta\ln L$ map is obtained, effectively, from convolution of the data with a kernel $\sim\ln(s/b+1)$ (cf.~eq.~\ref{eq:orig_lkl_ratio}), which has a width approximately equal to that of the PSF. This makes the $\Delta\ln L$ map smooth and its values not independent for close separations. The distribution of the local maxima peak heights in this case are most easily obtained through Monte Carlo simulations. 
The simulation results for a representative \art\ survey are shown in Fig.~\ref{fig:dist}b. Note that because the distribution of $\Delta \ln L$ is background-independent, we can expect that the distribution of peaks is also not very sensitive to the background level, and this is indeed the case (cf.\ black and red lines in Fig.~\ref{fig:dist}b).  

\citet{2018A&A...616A..25V} discuss the difficulty of analytical computation of the peak height distribution function in the noise field with non-Gaussian distribution, and outline an approximate approach based on the order statistics method \citep{Hogg2013}. In the limit of high peak values (or small probabilities of chance fluctuations), this approach gives the false alarm probability that equals the complimentary cumulative distribution function (CCDF) of the noise field, $\Phi(z)$, times the effective number of independent pixels in the image \citep[c.f.\ eq.17 in][]{2018A&A...616A..25V}, where $z$ is the peak height. In our case, $z=\Delta\ln L$ and $\Phi(z)$ is the CCDF for the $\chi^2_1$ distibution, $\Phi(z)=1-\Gamma(2z,1/2)$. This approach based on the order statistics is approximate, but we find that it can be modified as follows to very accurately describe the distribution of false detection observed in the Monte-Carlo simulations.

Let us consider a grid of pre-defined locations on the $\Delta\ln L$ map, separated by a distance larger than the size of the effective filter ($\approx$ the PSF size). The distribution of the maximal value on this grid can be precisely obtained with the order statistics approach. However, the true peak in the smoothed image will almost never coincide with the nearest grid point, and hence its value will be \emph{somewhat higher} than the maximum on the grid. The true peak value in the image is then the maximum on the grid plus an additional random variable whose distribution reflects the random offset between the true peak and grid points, and how fast the $\Delta\ln L$ map falls off from the points of its local maxima. The resulting distribution is a convolution of the probability density function given by the order statistics approach with a kernel skewed towards positive values. For a broad class of the CCDFs, this operation is approximately equivalent to simply shifting CCDF to the right by a constant log-offset, or, equivalently using $\Phi(kz)$ with $k\lesssim 1$. Indeed, we find that in our case, the function $1-\Gamma(0.89\times 2\Delta \ln L,1/2)$ provides an excellent description of the distribution of false detections in our Monte-Carlo simulations (dashed line in Fig.\ref{fig:dist}b). We will explore this technique further in a follow-up paper (Semena \& Vikhlinin, in preparation).

The stability with respect to the background is a very useful property as it allows us to greatly simplify the statistical calibration of the detection process in a survey.
Estimation of the number of false detection still requires a Monte Carlo simulation, but the simulations are needed for different PSF conditions and not for different levels of the background. In the \art\ survey case, there are multiple scans of any point in the sky and the resulting effective PSF is uniform. Therefore, a single set of simulations is required for statistical calibration of the survey.

The exact value of the $\Delta\ln L$ threshold is determined from a tradeoff between the desired low number of false detections and sensitivity. For \art\ surveys of a few tens of square degrees, a good choice is $\Delta \ln{L}_{\rm lim}=11.4$, resulting in 0.025 false sources per square degree (Fig.~\ref{fig:dist}). Note that for applications such as measuring the $\log N - \log S$ distribution of detected sources or their luminosity function, we need detection thresholds defined as a function of flux. Fortunately, a relationship between flux and $\Delta \ln L$ is well-defined (Fig.~\ref{fig:delta_ln_l_distribution}). 
We defer further discussion to a future work where we will present the $\log N - \log S$ of \art\ sources.

\subsection{Additional information from source spectra and grades}
\label{sec:energy_and_grades}

\begin{figure*}
    \centering
    \includegraphics[width=0.495\textwidth]{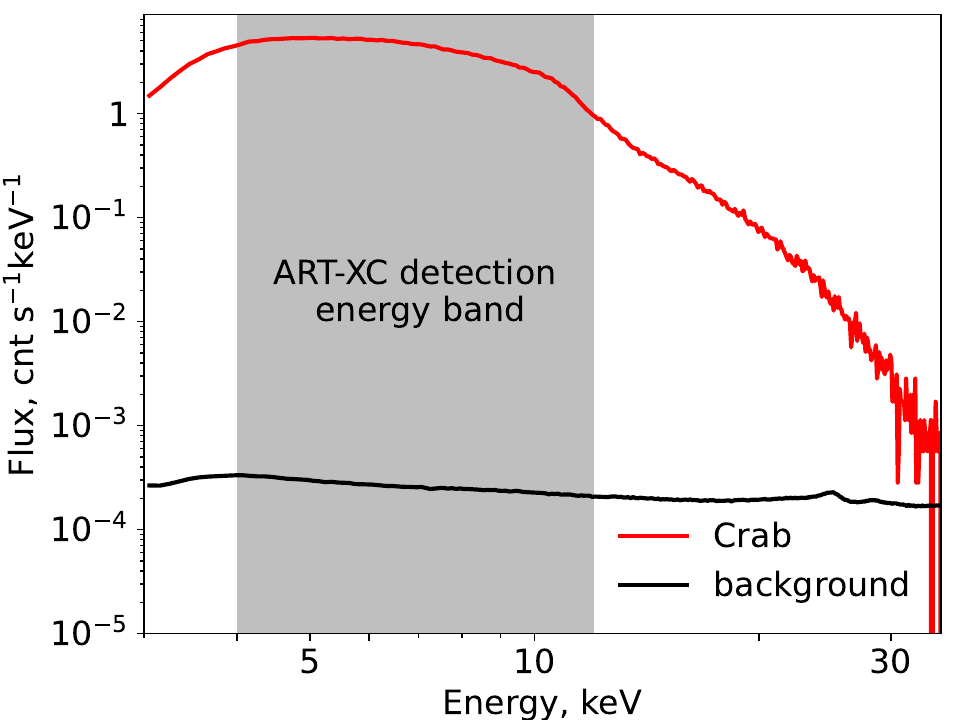}
    \includegraphics[width=0.495\textwidth]{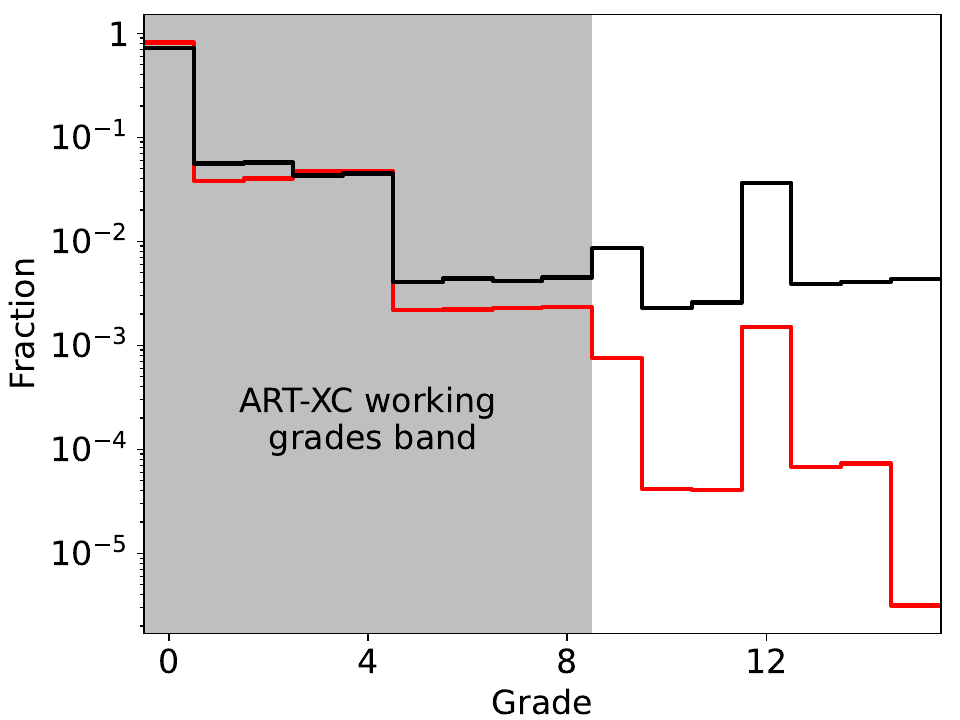}
    \caption{Differences between the \art\ source and background response. \emph{Left:} The typical source and background spectra. The background spectrum is normalized to the same extraction area ($2'$ circle). A drop in the source counts at $E>11.2$ is due to a sharp decrease in the mirror effective area. Therefore, an energy band of $4-12$~keV is typically used for source detection. \emph{Right:} Distribution of the event grades for the source and background signals. The grades $>9$ are routinely filtered out because they are virtually absent in the source signal. Notice that there are non-negligible differences between the source and background response within the detection energy band and the working grade range. These differences can be exploited to increase the source detection sensitivity (see text).}
    \label{fig:energy_and_grade_distribution}
\end{figure*}

So far, we have considered a case of source detection in a single (broad) energy band. In principle, there is additional information that can be used to improve the detection sensitivity. For example, the source spectra are typically very distinct from that of the background (see Fig.\ref{fig:energy_and_grade_distribution}). This difference can be incorporated into the likelihood model if one is aiming to detect sources of a specific spectral type. In addition, the background and real source events typically have a different detector grade distribution \citep[see  \S\ref{ap:grades} and description of the \art\ detectors in ][]{2014SPIE.9144E..13L}. Grades can significantly help with rejecting the background in X-ray detectors \citep[see, e.g.][]{1983NIMPR.213..201B, 1988ITNS...35..534L}. For \art\ case, the relative frequency of different grades from the background and from real photons from the on-board calibration source are shown in the right panel of Fig.\ref{fig:energy_and_grade_distribution}. Events with $\mathrm{grade}>8$ are dominated by the background and they are by default excluded from the analysis. However, there are non-negligible differences between the source and background grade distribution for $\mathrm{grade}\le8$, and this can also be incorporated in the likelihood model. Specifically, we can replace $s$ and $b$ in eqs.~\ref{eq:source:model} and~\ref{eq:poisson_prob} with
\begin{equation}
\label{eq:source:model:E:grade}
\begin{split}
s(\xi,\eta,E,G) = f(E)\; p_S(G,E) \; V(\xi,\eta,E)\; \times \\ \int \mathit{PSF}(x-\xi,y-\eta,\alpha,\beta,E)\,dx\,dy,
\end{split}
\end{equation}
$E$ is the registered photon energy, $G$ is the event grade, $f(E)$ is a normalized source spectrum convolved with the telescope response, $p_S(G,E)$ is the detector grade distribution for ``sky'' events, $V(\xi,\eta,E)$ and $\mathit{PSF}(\Delta x,\Delta y,\alpha,\beta,E)$ are the energy dependent vignetting and PSF, respectively. For the background model, 
\begin{equation}
\label{eq:background:model:E:grade} 
b_i \; \longrightarrow \; b_i(E,G) = B\,f_B(E),p_B(G,E),
\end{equation}
where $f_B(E)$ is the normalized background spectrum, $p_B(G,E)$ is the detector grade distribution for background events, and $B$ is the overall background intensity. For a source with a Crab-like spectrum, 
using the spectral and grade information for source detection leads to an increase of $\Delta \ln L\approx 0.6$, which is equivalent to an additional $\approx 1\sigma$ signal being added to the source detection process. 

\subsection{Source parameter uncertainties}
\label{sec:confidece_intervals}
The value of R that corresponds to the local and the maximum's location are automatically the maximum likelihood estimation of the source flux and position respectively. 
A small additional analysis is required to derive uncertainties on the source parameters, such as flux and position. They can be obtained, e.g., with a Markov chain sampling technique \citep[we use an implementation described in][]{2015arXiv150708050S}. This is a natural choice in our case because the entire detection approach is based on computing the likelihood (probabilities) given the source parameters (eq.\ref{eq:poisson_prob}). 
Alternatively, one can use the \cite{1979ApJ...228..939C} approach based on varying the source parameters around the best-fit value until $\ln L$ decreases by a specific value; for a example a change of $\ln L$ by $-0.5$ corresponds to the boundaries of a single-parameter 68\% confidence interval. We note that based on this approach, the source position uncertainties can be obtained directly from the likelihood ratio map we use for source detection: a region around a local maximum within $-1.15$ of the peak value\footnote{Equivalent to $\Delta \chi^2=2.3$ for a two-parameter 68\% confidence region.} corresponds to the 68\% CL uncertainty of the source position. We have verified via bootstrap approach that this method does result in the correct positional uncertainties.

\section{ART-XC Specific Issues}

Detection method outlined about is generic and can be applied to any X-ray imaging telescope. There are always however instrument-specific issues. In this section, we describe a treatment of two such issues in the \art\ case. 

\subsection{Illumination from bright sources}
\label{sec:illumination}

The ART-XC PSF has strong wings, resulting in ``halos'' around bright sources extending to a distance of up to 1 degree. These halos are formed due to single reflections in the mirror system of the telescope \cite[see][]{Nariai:88, 2017SPIE10399E..0JB, 2019ExA....48..233P, 2020JInst..15P1032B}. They are notoriously hard to model \citep[but see e.g.][]{2015A&A...573A..22R}.

\begin{figure*}
\centerline{\includegraphics[width=0.9\textwidth]{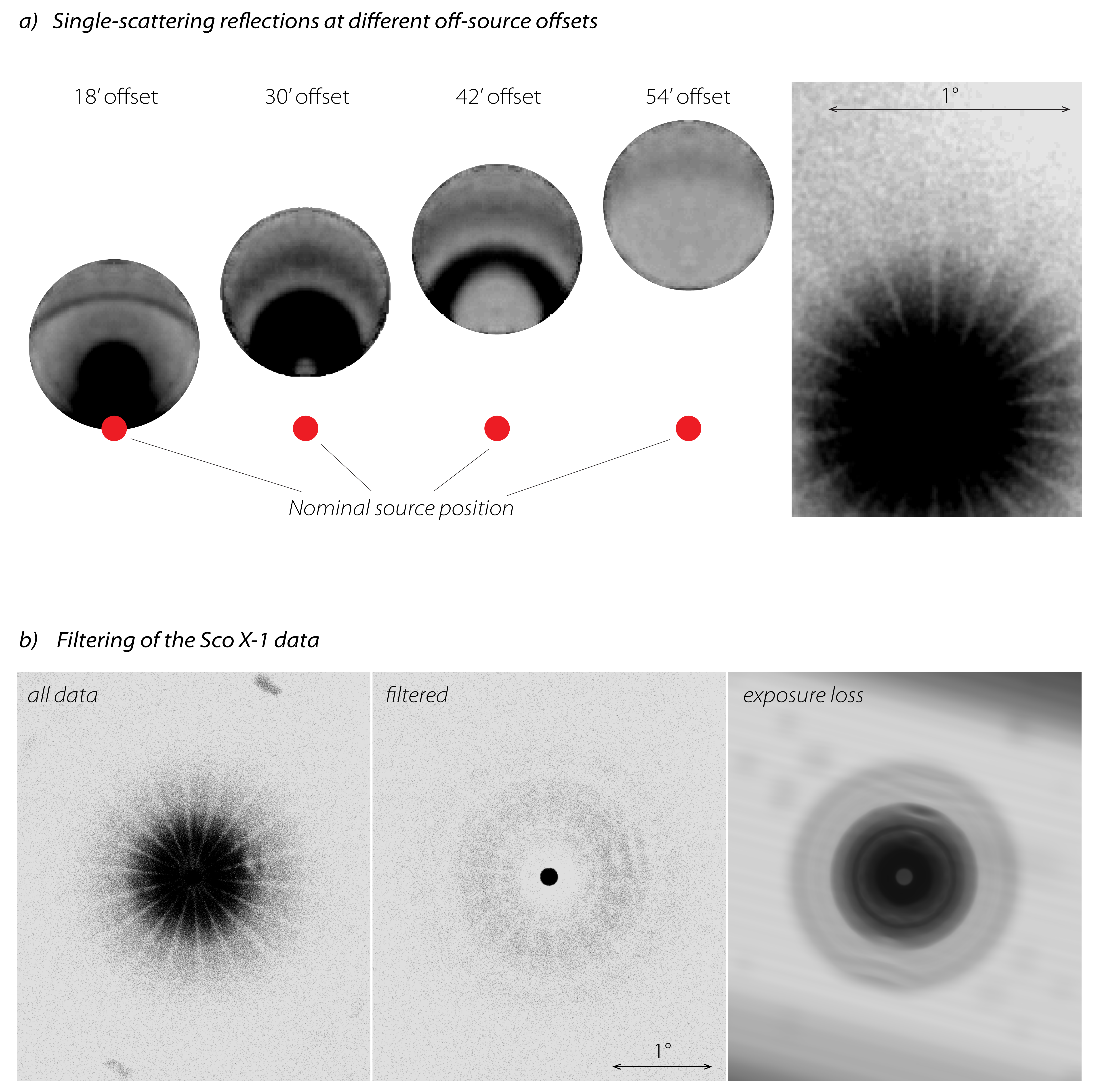}}
    \caption{Single reflections in the telescope mirror system result in large, degree-scale halos around bright sources. However, the effect can be suppressed through masking of specific areas on the detector. \emph{Top:} Single reflections halos at off-source offsets $18'$ (i.e., with the source exactly at the edge of the \art\ field of view), $30'$, $42'$, and $54'$, in comparison with the accumulated halo around Sco-X1 during the all-sky survey. Even though the integrated halo extends to $\sim 1^\circ$ radius, part of the field of view is little affected by single reflections at offsets as small as $10'$. Therefore, the appropriate filtering masks can be created and applied to the data to reduce the effect of scattering. \emph{Bottom:}
    Example of the single reflections halo filtering using the Sco X-1 all-sky survey data. The \emph{left} panel shows the Sco X-1 image with no filtering applied. The \emph{middle} panel shows the result of filtering out the most affected areas at each off-source position. The $1^\circ$ single-bounce scatter is drastically suppressed. This is done at an expense of reducing the effective exposure by $\sim 50\%$ over most of the halo-affected area, even though the loss is $95\%$ in the immediate vicinity of the source (\emph{right panel}).}
    \label{fig:sco_illumination_reduction}
\end{figure*}

For a fixed position of the source relative to the optical axis of the mirror module, singly reflected photons occupy a limited area in the detector plane. During the scan, the affected areas move and change morphology with a speed different from that of the scan (see top panels in Fig.\ref{fig:sco_illumination_reduction}). At any given moment in time we can predict where the most affected areas are and mask those areas from further analysis. This approach strongly suppresses the effect of scattering halos on detection of other source and at the same time does not lead to a complete loss of data at any sky position. 

Based on observations of bright sources during the calibration phase, we created a library of illumination masks from bright point sources for a set of offsets from the telescope optical axis, ranging from 5 to 85 arcmin. The masks were created such that the illumination is suppressed by a factor of $15$ at distances in a range of $15-50\arcmin$ from the bright source (see fig.\ref{fig:sco_illumination_reduction}). Such filtering results in a $\sim 50\%$ exposure loss in a $\sim 30'$ region around bright sources  (Fig.\ref{fig:sco_illumination_reduction}).

\subsection{Background model}
\label{sec:background_model}

The ART-XC background consists of three main components: the detector particle-induced background, diffuse radiation from the sky, and the residual halo illumination from bright sources (\S\ref{sec:illumination}). Except for the regions of bright diffuse emission in the Galactic Center region and the Galactic ridge, by far the dominant background component is the particle-induced one ($\sim 99.5\%$ of the total). It is very stable \citep[except for short episodes of solar flares, see][]{art_xc_allsky_survey} and had median value of $2\times10^{-4}$~s$^{-1}$ per pixel in 4--12~keV energy band during the first two years of operation.  Large background data sets were accumulated during ART-XC scanning observations in the source-free regions. With a small renormalization using the observed background intensity above $\sim 40$ keV\footnote{At these energies, the mirror effective area is small and thus there is no astrophysical background contribution.}, these datasets can be used to accurately predict the particle-induced background at all locations on the detector and throughout the useful energy band. The renormalization accuracy is better than $1\%$ for time intervals of $1\,$ksec, which is much shorter than the typical time scale of the secular changes in the particle-induced background.

The second component we need to model is the astrophysical diffuse emission. In practice, it is significant only in the Galactic Center and Ridge regions. These are also the regions densely populated with bright point sources. Despite the application of the illumination suppression algorithm (see the \ref{sec:illumination} paragraph), a certain fraction of their extended halo remains. A typical angular scale of this residual emission is $\sim 1$\deg. We do not distinguish these residual halos from the truly diffuse Galactic emission or from emission from a large number of point sources well below the ART-XC detection threshold \citep[c.f.,][]{2009Natur.458.1142R}. The diffuse sky background model for ART-XC is derived as follows. 

First, we subtract the particle-induced background model from the observed photon map. This map is flat-fielded using the exposure map including mirror vignetting. 
Then we remove small-scale features associated with detectable sources using a wavelet decomposition method with an implementation similar to that described by \cite{1998ApJ...502..558V}. With wavelet decomposition we remove all detectable structures on spatial scales of 20, 40, 80 and 160\arcsec\ are discarded, i.e. comparable to the width of the ART-XC PSF. The residual map is smoothed with a Gaussian filter with a width of 10\arcmin. 
This smoothing scale was chosen because it sufficiently suppresses the photon counting statics noise, while at the same time remains small compared to the typical scale of the Galactic background. The resulting map gives an estimate of the diffuse background count rate from a given region in the sky, which is used --- together with the particle-induced background map ---  in the likelihood ratio calculations.

\section{Summary and Conclusions}

\begin{figure*}

    \centerline{\includegraphics[width=0.8\textwidth]{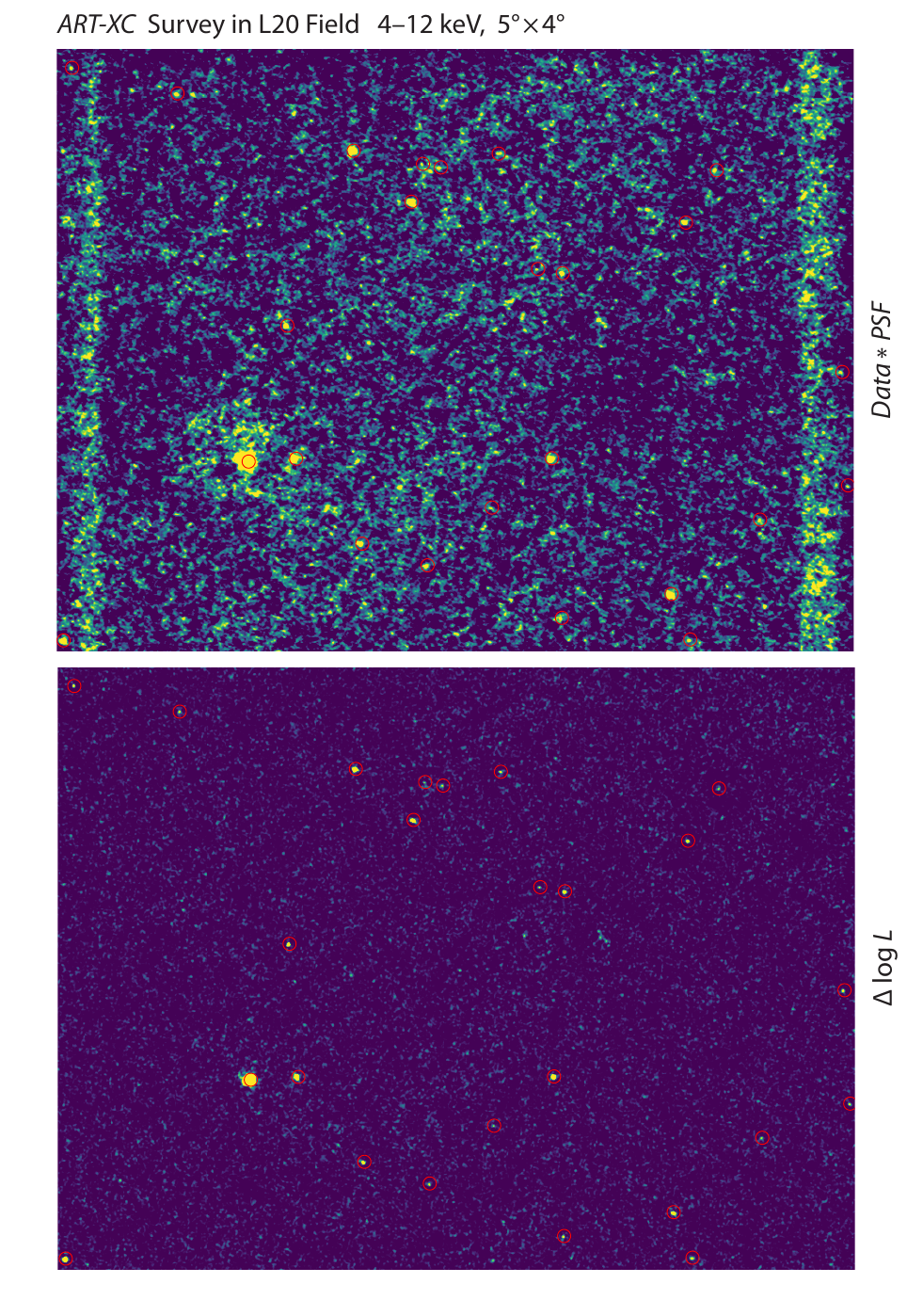}}
    \caption{The application of our source detection algorithm to the \art\ 'L20' field. \emph{Top:} the total photon image convolved with the average PSF, as would be used for traditional source detection. Notice significant exposure and background non-uniformities across the field. \emph{Bottom:} the $\Delta \ln{L}$ map in the same region, with the color map normalized such that the sources have the same brightness in both maps. Red circles mark the location of sources detected above the $\Delta\ln{L}=11.4$ threshold. Notice the great reduction in the ``noise'' level in the $\Delta\ln{L}$, the uniformity of noise, and the high-confidence detection of the sources. These improvements are due to the maximally optimal use of available information.}
      \label{fig:L20}
\end{figure*}

Putting all components of our source detection pipeline together, we show in Figure~\ref{fig:L20} the application of our algorithm to the data in the \art\ 'L20' field (a $5\deg\times4\deg$ field in the Galactic Plane region centered at $l=20\deg$, $b=0\deg$, also check \citealt{GC_article} (Paper II), for the analysis of Galactic Center field, populated with several bright X-ray sources). The top panel shows the total photon image convolved with the field of view averaged PSF. This is the type of data used in the traditional source detection pipelines. Notice significant exposure and background non-uniformities across the field. As a result, the surface brightness peaks at the location of detected sources (marked by red circles) are not very prominent. Obviously, a confident source detection using such data is possible only at a very high level of detection threshold, and there will be significant variations of the threshold value across the field. 

The situation is drastically better in the $\Delta\ln{L}$ map (\S\ref{sec:likelihood_ratio}), in the sense that the fluctuations in the source-free regions in this map are uniform (as expected, \S\ref{sec:detection_threshold}) and have a much lower amplitude  compared to the peaks corresponding to sources. Twenty five sources are clearly detected above a threshold of $\Delta \ln{L}>11.4$. Less than one false detection is expected in $20$~deg$^2$ for this choice of threshold (\S\ref{sec:detection_threshold}), so most if not all of these detections are real sources (Karasev et al., in preparation). For comparison, the PSF convolution algorithm produces $110$ false detection for the same flux limit. Alternatively, if we set the threshold for the PSF convolution such that there is only one false detection, then only 14 of 25 sources are detected. This comparison shows that for the ART-XC data, we are far from the high-background regime where the likelihood-based approach and PSF convolution become equivalent. The improvements in sensitivity of the source detection using the $\Delta\ln{L}$ compared with the PSF-smoothed image are due to the maximally optimal use of available information (\S\ref{sec:likelihood_ratio}). Primarily, the main improvement is due to using separately the data with the sharp PSF on-axis and those with the degraded PSF off-axis. Additional small improvements in sensitivity are due to using photon energies and grades (\S\ref{sec:energy_and_grades}).

To conclude, we presented an X-ray source detection method entirely based on the likelihood analysis and demonstrated the performance using the \art\ data. Our method results in uniform detection conditions even using very non-uniform data (PSF, background level). This stability greatly reduces the effort required for further statistical calibration needed, e.g., to obtain the $\log N - \log S$ or the source luminosity function. We will present this analysis in application to \art\ in a future paper. In the next paper in this series \citep{GC_article}, we present a catalog of sources detected using this method in a deep Galactic Bulge survey with \art.

\section*{Acknowledgements}

This work is based on observations with  Mikhail Pavlinsky \art\ telescope, hard X-ray instrument on board the \srg\  observatory. The \srg\ observatory was created by Roskosmos (the Lavochkin Association and its subcontractors) in the interests of the Russian Academy of Sciences represented by its Space Research Institute (IKI) in the framework of the Russian Federal Space Program, with the participation of Germany. The \art\ team thanks the Russian Space Agency, Russian Academy of Sciences and State Corporation Rosatom for the support of the  \art\ telescope design and development. The science data are downlinked via the Deep Space Network Antennae in Bear Lakes, Ussurijsk, and Baykonur, funded by Roskosmos. AV's work on this project was partially supported by the CXO grant AR1-22013X.

\bibliographystyle{mnras}
\bibliography{biblio.bib}

\begin{thebibliography}{}
\makeatletter
\relax
\def\mn@urlcharsother{\let\do\@makeother \do\$\do\&\do\#\do\^\do\_\do\%\do\~}
\def\mn@doi{\begingroup\mn@urlcharsother \@ifnextchar [ {\mn@doi@}
  {\mn@doi@[]}}
\def\mn@doi@[#1]#2{\def\@tempa{#1}\ifx\@tempa\@empty \href
  {http://dx.doi.org/#2} {doi:#2}\else \href {http://dx.doi.org/#2} {#1}\fi
  \endgroup}
\def\mn@eprint#1#2{\mn@eprint@#1:#2::\@nil}
\def\mn@eprint@arXiv#1{\href {http://arxiv.org/abs/#1} {{\tt arXiv:#1}}}
\def\mn@eprint@dblp#1{\href {http://dblp.uni-trier.de/rec/bibtex/#1.xml}
  {dblp:#1}}
\def\mn@eprint@#1:#2:#3:#4\@nil{\def\@tempa {#1}\def\@tempb {#2}\def\@tempc
  {#3}\ifx \@tempc \@empty \let \@tempc \@tempb \let \@tempb \@tempa \fi \ifx
  \@tempb \@empty \def\@tempb {arXiv}\fi \@ifundefined
  {mn@eprint@\@tempb}{\@tempb:\@tempc}{\expandafter \expandafter \csname
  mn@eprint@\@tempb\endcsname \expandafter{\@tempc}}}

\bibitem[\protect\citeauthoryear{{Abdo} et~al.,}{{Abdo}
  et~al.}{2010}]{2010ApJS..188..405A}
{Abdo} A.~A.,  et~al., 2010, \mn@doi [\apjs] {10.1088/0067-0049/188/2/405},
  \href {https://ui.adsabs.harvard.edu/abs/2010ApJS..188..405A} {188, 405}

\bibitem[\protect\citeauthoryear{{Bailey}, {Damerell}, {English}, {Gillman},
  {Lintern}, {Watts}  \& {Wickens}}{{Bailey}
  et~al.}{1983}]{1983NIMPR.213..201B}
{Bailey} R.,  {Damerell} C.~J.~S.,  {English} R.~L.,  {Gillman} A.~R.,
  {Lintern} A.~L.,  {Watts} S.~J.,   {Wickens} F.~J.,  1983, \mn@doi [Nuclear
  Instruments and Methods in Physics Research] {10.1016/0167-5087(83)90413-1},
  \href {https://ui.adsabs.harvard.edu/abs/1983NIMPR.213..201B} {213, 201}

\bibitem[\protect\citeauthoryear{{Brunner} et~al.,}{{Brunner}
  et~al.}{2022}]{2022A&A...661A...1B}
{Brunner} H.,  et~al., 2022, \mn@doi [\aap] {10.1051/0004-6361/202141266},
  \href {https://ui.adsabs.harvard.edu/abs/2022A&A...661A...1B} {661, A1}

\bibitem[\protect\citeauthoryear{{Buitrago-Casas} et~al.,}{{Buitrago-Casas}
  et~al.}{2017}]{2017SPIE10399E..0JB}
{Buitrago-Casas} J.~C.,  et~al., 2017, in {O'Dell} S.~L.,  {Pareschi} G.,  eds,
   Society of Photo-Optical Instrumentation Engineers (SPIE) Conference Series
  Vol. 10399, Society of Photo-Optical Instrumentation Engineers (SPIE)
  Conference Series. p. 103990J, \mn@doi{10.1117/12.2274675}

\bibitem[\protect\citeauthoryear{{Buitrago-Casas} et~al.,}{{Buitrago-Casas}
  et~al.}{2020}]{2020JInst..15P1032B}
{Buitrago-Casas} J.~C.,  et~al., 2020, \mn@doi [Journal of Instrumentation]
  {10.1088/1748-0221/15/11/P11032}, \href
  {https://ui.adsabs.harvard.edu/abs/2020JInst..15P1032B} {15, P11032}

\bibitem[\protect\citeauthoryear{{Cash}}{{Cash}}{1979}]{1979ApJ...228..939C}
{Cash} W.,  1979, \mn@doi [\apj] {10.1086/156922}, \href
  {https://ui.adsabs.harvard.edu/abs/1979ApJ...228..939C} {228, 939}

\bibitem[\protect\citeauthoryear{{Cruddace}, {Hasinger}  \&
  {Schmitt}}{{Cruddace} et~al.}{1988}]{1988ESOC...28..177C}
{Cruddace} R.~G.,  {Hasinger} G.~H.,   {Schmitt} J.~H.,  1988, in European
  Southern Observatory Conference and Workshop Proceedings. pp 177--182

\bibitem[\protect\citeauthoryear{{Ehlert} et~al.,}{{Ehlert}
  et~al.}{2022}]{2022MNRAS.515.5185E}
{Ehlert} S.,  et~al., 2022, \mn@doi [\mnras] {10.1093/mnras/stac2072}, \href
  {https://ui.adsabs.harvard.edu/abs/2022MNRAS.515.5185E} {515, 5185}

\bibitem[\protect\citeauthoryear{{Evans} et~al.,}{{Evans}
  et~al.}{2010}]{2010ApJS..189...37E}
{Evans} I.~N.,  et~al., 2010, \mn@doi [\apjs] {10.1088/0067-0049/189/1/37},
  \href {https://ui.adsabs.harvard.edu/abs/2010ApJS..189...37E} {189, 37}

\bibitem[\protect\citeauthoryear{{Evans} et~al.,}{{Evans}
  et~al.}{2020}]{2020ApJS..247...54E}
{Evans} P.~A.,  et~al., 2020, \mn@doi [\apjs] {10.3847/1538-4365/ab7db9}, \href
  {https://ui.adsabs.harvard.edu/abs/2020ApJS..247...54E} {247, 54}

\bibitem[\protect\citeauthoryear{{Gabriel} et~al.,}{{Gabriel}
  et~al.}{2004}]{2004ASPC..314..759G}
{Gabriel} C.,  et~al., 2004, in {Ochsenbein} F.,  {Allen} M.~G.,   {Egret} D.,
  eds,  Astronomical Society of the Pacific Conference Series Vol. 314,
  Astronomical Data Analysis Software and Systems (ADASS) XIII. p.~759

\bibitem[\protect\citeauthoryear{{Gilli}, {Comastri}  \& {Hasinger}}{{Gilli}
  et~al.}{2007}]{2007A&A...463...79G}
{Gilli} R.,  {Comastri} A.,   {Hasinger} G.,  2007, \mn@doi [\aap]
  {10.1051/0004-6361:20066334}, \href
  {https://ui.adsabs.harvard.edu/abs/2007A&A...463...79G} {463, 79}

\bibitem[\protect\citeauthoryear{{Hands}, {Warwick}, {Watson}  \&
  {Helfand}}{{Hands} et~al.}{2004}]{2004MNRAS.351...31H}
{Hands} A.~D.~P.,  {Warwick} R.~S.,  {Watson} M.~G.,   {Helfand} D.~J.,  2004,
  \mn@doi [\mnras] {10.1111/j.1365-2966.2004.07777.x}, \href
  {https://ui.adsabs.harvard.edu/abs/2004MNRAS.351...31H} {351, 31}

\bibitem[\protect\citeauthoryear{{Hartman} et~al.,}{{Hartman}
  et~al.}{1999}]{1999ApJS..123...79H}
{Hartman} R.~C.,  et~al., 1999, \mn@doi [\apjs] {10.1086/313231}, \href
  {https://ui.adsabs.harvard.edu/abs/1999ApJS..123...79H} {123, 79}

\bibitem[\protect\citeauthoryear{{Helstrom}}{{Helstrom}}{1968}]{1968IJTP....1...37H}
{Helstrom} C.~W.,  1968, \mn@doi [International Journal of Theoretical Physics]
  {10.1007/BF00668829}, \href
  {https://ui.adsabs.harvard.edu/abs/1968IJTP....1...37H} {1, 37}

\bibitem[\protect\citeauthoryear{{Hogg}, {McKean}  \& {Craig}}{{Hogg}
  et~al.}{2013}]{Hogg2013}
{Hogg} R.~V.,  {McKean} J.~W.,   {Craig} A.~T.,  2013, {Introduction to
  Mathematical Statistics}.
New York: Pearson

\bibitem[\protect\citeauthoryear{{Krivonos} et~al.,}{{Krivonos}
  et~al.}{2017}]{2017ExA....44..147K}
{Krivonos} R.,  et~al., 2017, \mn@doi [Experimental Astronomy]
  {10.1007/s10686-017-9555-0}, \href
  {https://ui.adsabs.harvard.edu/abs/2017ExA....44..147K} {44, 147}

\bibitem[\protect\citeauthoryear{Lehmann \& Romano}{Lehmann \&
  Romano}{2005}]{lehmann2005testing}
Lehmann E.~L.,  Romano J.~P.,  2005, Testing statistical hypotheses, third edn.
Springer Texts in Statistics, Springer, New York

\bibitem[\protect\citeauthoryear{{Levin}, {Pavlinsky}, {Akimov}, {Kuznetsova},
  {Rotin}, {Krivchenko}, {Lapshov}  \& {Oleinikov}}{{Levin}
  et~al.}{2014}]{2014SPIE.9144E..13L}
{Levin} V.,  {Pavlinsky} M.,  {Akimov} V.,  {Kuznetsova} M.,  {Rotin} A.,
  {Krivchenko} A.,  {Lapshov} I.,   {Oleinikov} V.,  2014, in {Takahashi} T.,
  {den Herder} J.-W.~A.,   {Bautz} M.,  eds,  Society of Photo-Optical
  Instrumentation Engineers (SPIE) Conference Series Vol. 9144, Space
  Telescopes and Instrumentation 2014: Ultraviolet to Gamma Ray. p. 914413,
  \mn@doi{10.1117/12.2056311}

\bibitem[\protect\citeauthoryear{{Lumb} \& {Holland}}{{Lumb} \&
  {Holland}}{1988}]{1988ITNS...35..534L}
{Lumb} D.~H.,  {Holland} A.~D.,  1988, \mn@doi [IEEE Transactions on Nuclear
  Science] {10.1109/23.12780}, \href
  {https://ui.adsabs.harvard.edu/abs/1988ITNS...35..534L} {35, 534}

\bibitem[\protect\citeauthoryear{{Lynx Team}}{{Lynx Team}}{2019}]{LYNXREPORT}
{Lynx Team} 2019, in \emph{Lynx} X-Ray Observatory Concept Study Report,
  \url{https://www.lynxobservatory.org}.

\bibitem[\protect\citeauthoryear{{Masias}, {Freixenet}, {Llad{\'o}}  \&
  {Peracaula}}{{Masias} et~al.}{2012}]{2012MNRAS.422.1674M}
{Masias} M.,  {Freixenet} J.,  {Llad{\'o}} X.,   {Peracaula} M.,  2012, \mn@doi
  [\mnras] {10.1111/j.1365-2966.2012.20742.x}, \href
  {https://ui.adsabs.harvard.edu/abs/2012MNRAS.422.1674M} {422, 1674}

\bibitem[\protect\citeauthoryear{{Mattox} et~al.,}{{Mattox}
  et~al.}{1996}]{1996ApJ...461..396M}
{Mattox} J.~R.,  et~al., 1996, \mn@doi [\apj] {10.1086/177068}, \href
  {https://ui.adsabs.harvard.edu/abs/1996ApJ...461..396M} {461, 396}

\bibitem[\protect\citeauthoryear{Nariai}{Nariai}{1988}]{Nariai:88}
Nariai K.,  1988, \mn@doi [Appl. Opt.] {10.1364/AO.27.000345}, 27, 345

\bibitem[\protect\citeauthoryear{{Neyman} \& {Pearson}}{{Neyman} \&
  {Pearson}}{1933}]{1933RSPTA.231..289N}
{Neyman} J.,  {Pearson} E.~S.,  1933, \mn@doi [Philosophical Transactions of
  the Royal Society of London Series A] {10.1098/rsta.1933.0009}, \href
  {https://ui.adsabs.harvard.edu/abs/1933RSPTA.231..289N} {231, 289}

\bibitem[\protect\citeauthoryear{{Ofek} \& {Zackay}}{{Ofek} \&
  {Zackay}}{2018}]{2018AJ....155..169O}
{Ofek} E.~O.,  {Zackay} B.,  2018, \mn@doi [\aj] {10.3847/1538-3881/aab265},
  \href {https://ui.adsabs.harvard.edu/abs/2018AJ....155..169O} {155, 169}

\bibitem[\protect\citeauthoryear{{Pavlinsky} et~al.,}{{Pavlinsky}
  et~al.}{2019}]{2019ExA....48..233P}
{Pavlinsky} M.,  et~al., 2019, \mn@doi [Experimental Astronomy]
  {10.1007/s10686-019-09646-8}, \href
  {https://ui.adsabs.harvard.edu/abs/2019ExA....48..233P} {48, 233}

\bibitem[\protect\citeauthoryear{{Pavlinsky} et~al.,}{{Pavlinsky}
  et~al.}{2021}]{2021A&A...650A..42P}
{Pavlinsky} M.,  et~al., 2021, \mn@doi [\aap] {10.1051/0004-6361/202040265},
  \href {https://ui.adsabs.harvard.edu/abs/2021A&A...650A..42P} {650, A42}

\bibitem[\protect\citeauthoryear{{Pavlinsky} et~al.,}{{Pavlinsky}
  et~al.}{2022}]{art_xc_allsky_survey}
{Pavlinsky} M.,  et~al., 2022, \mn@doi [\aap] {10.1051/0004-6361/202141770},
  \href {https://ui.adsabs.harvard.edu/abs/2022A&A...661A..38P} {661, A38}

\bibitem[\protect\citeauthoryear{{Raimondi} \& {Spiga}}{{Raimondi} \&
  {Spiga}}{2015}]{2015A&A...573A..22R}
{Raimondi} L.,  {Spiga} D.,  2015, \mn@doi [\aap]
  {10.1051/0004-6361/201424907}, \href
  {https://ui.adsabs.harvard.edu/abs/2015A&A...573A..22R} {573, A22}

\bibitem[\protect\citeauthoryear{{Revnivtsev}, {Sazonov}, {Churazov}, {Forman},
  {Vikhlinin}  \& {Sunyaev}}{{Revnivtsev} et~al.}{2009}]{2009Natur.458.1142R}
{Revnivtsev} M.,  {Sazonov} S.,  {Churazov} E.,  {Forman} W.,  {Vikhlinin} A.,
   {Sunyaev} R.,  2009, \mn@doi [\nat] {10.1038/nature07946}, \href
  {https://ui.adsabs.harvard.edu/abs/2009Natur.458.1142R} {458, 1142}

\bibitem[\protect\citeauthoryear{Rios \& Sahinidis}{Rios \&
  Sahinidis}{2013}]{zbMATH06193572}
Rios L.~M.,  Sahinidis N.~V.,  2013, \mn@doi [J. Glob. Optim.]
  {10.1007/s10898-012-9951-y}, 56, 1247

\bibitem[\protect\citeauthoryear{{Salvatier}, {Wiecki}  \&
  {Fonnesbeck}}{{Salvatier} et~al.}{2015}]{2015arXiv150708050S}
{Salvatier} J.,  {Wiecki} T.,   {Fonnesbeck} C.,  2015, arXiv e-prints, \href
  {https://ui.adsabs.harvard.edu/abs/2015arXiv150708050S} {p. arXiv:1507.08050}

\bibitem[\protect\citeauthoryear{{Semena} et~al.,}{{Semena}
  et~al.}{2023}]{GC_article}
{Semena} A.,  et~al., 2023, submitted to MNRAS, Paper II

\bibitem[\protect\citeauthoryear{{Sunyaev} et~al.,}{{Sunyaev}
  et~al.}{2021}]{srg}
{Sunyaev} R.,  et~al., 2021, \mn@doi [\aap] {10.1051/0004-6361/202141179},
  \href {https://ui.adsabs.harvard.edu/abs/2021A&A...656A.132S} {656, A132}

\bibitem[\protect\citeauthoryear{{Vikhlinin}, {McNamara}, {Forman}, {Jones},
  {Quintana}  \& {Hornstrup}}{{Vikhlinin} et~al.}{1998}]{1998ApJ...502..558V}
{Vikhlinin} A.,  {McNamara} B.~R.,  {Forman} W.,  {Jones} C.,  {Quintana} H.,
  {Hornstrup} A.,  1998, \mn@doi [\apj] {10.1086/305951}, \href
  {https://ui.adsabs.harvard.edu/abs/1998ApJ...502..558V} {502, 558}

\bibitem[\protect\citeauthoryear{{Vio} \& {Andreani}}{{Vio} \&
  {Andreani}}{2018}]{2018A&A...616A..25V}
{Vio} R.,  {Andreani} P.,  2018, \mn@doi [\aap] {10.1051/0004-6361/201832641},
  \href {https://ui.adsabs.harvard.edu/abs/2018A&A...616A..25V} {616, A25}

\bibitem[\protect\citeauthoryear{{Voges} et~al.,}{{Voges}
  et~al.}{1999}]{1999A&A...349..389V}
{Voges} W.,  et~al., 1999, \mn@doi [\aap] {10.48550/arXiv.astro-ph/9909315},
  \href {https://ui.adsabs.harvard.edu/abs/1999A&A...349..389V} {349, 389}

\bibitem[\protect\citeauthoryear{Wilks}{Wilks}{1938}]{10.1214/aoms/1177732360}
Wilks S.~S.,  1938, \mn@doi [The Annals of Mathematical Statistics]
  {10.1214/aoms/1177732360}, 9, 60

\bibitem[\protect\citeauthoryear{{Wolter}}{{Wolter}}{1952}]{1952AnP...445...94W}
{Wolter} H.,  1952, \mn@doi [Annalen der Physik] {10.1002/andp.19524450108},
  \href {https://ui.adsabs.harvard.edu/abs/1952AnP...445...94W} {445, 94}

\makeatother
\end{thebibliography}

\appendix
\section{\art\ grades}
\label{ap:grades}
The \art\ possess semiconductor double-sided strip detectors, which have their own specifics in events grades determination. 
Each event is stored in the telemetry with six amplitudes from three top and three bottom strips centered at event. 
The grade is determined by the comparison of these amplitudes to the specified signal thresholds. 
This provides us with sixteen independent grades, which could be classified as single (grade = 0), double (i.e. with two adjacent stripes in one or both surfaces, 0<grade<8) or multiple (grade>8). Multiple events are nearly impossible to produce with X-ray photon, therefore they are dominated by cosmic ray hits.

\bsp    
\label{lastpage}
 
\end{document}